B.M. Ovchinnikov[1], V.V. Parusov[1], Yu.B. Ovchinnikov[2]

# Multi-channel gas electron multiplier with metallic electrodes

[1]Institute of Nuclear Research of Russian Academy of Sciences, Troitsk, Russia
[2]National Physical Laboratory, Teddington, Middlesex, TW11 0LW, UK

## Abstract

The design of multi-channel gas electron multiplier (MGEM) with metallic electrodes is proposed, produced and tested. The electrodes of MGEM are produced from the brass plates with thickness of 1 mm, round openings of 1 mm in diameter and 1.5 mm steps between them. The gap between the electrodes is equal to 3 mm, while the total working area has a diameter of 20 mm.

The neon gas fillings of the MWGEM chamber with micro admixtures of $N_2$ and $H_2O$ have been tested.

The total maximal coefficient of proportional multiplication of electrons in neon with admixture of $(H_2O+N_2)\leq 100$ *ppm* of $3\cdot 10^4$ is obtained.

## Introduction

The multi-channel wire gas electron multipliers with gaps of 1 and 3 mm have been investigated in works [1-3]. These MWGEM, filled with commercial neon gas, have high reliability and rather high coefficients of amplification of $K_{ampl}^{max} \cong 10^5$-$10^6$, without streamers in proportional region.

One of the problems of the MWGEM is noise caused by the microphonic effect, which takes place in presence of mechanical vibrations.

To increase the mechanical durability of a gas electron multiplier and to suppress the microphonic effect in it, we propose the MGEM with metallic electrodes.

During operation of a MGEM, the air adsorbed on the walls of the chamber can be desorbed and admixed to the neon gas of the MGEM. For this reason it is necessary to investigate the influence of $H_2O$ and $N_2$ admixtures in neon to the operation of a MGEM.

## The design of MGEM and investigation of its properties

The electrodes of MGEM are produced from the brass plates with thickness of 1 mm, with round openings of 1 mm in diameter and 1.5 mm steps between them. The gap between the electrodes is equal to 3 mm, while the total working area has a diameter of 20 mm (Fig. 1).

The tests of MGEM were conducted in a chamber containing a cathode, MGEM and anode (Fig. 2). The gap between the cathode and MGEM is equal to 13 mm and a gap between the MGEM and the anode is equal to 6 mm.

The signals from the anode were amplified by a charge sensitive amplifier BUS 2-96.

The results of MGEM testing in a chamber, which was not preliminary baked for removal of air and water contaminations from its walls, are shown in Fig. 3, 4. The chamber was filled with a commercial neon gas, which is usually contaminated with $H_2O$, $N_2$ and $O_2$ at the level of about 2 ppm. On the other hand, the concentration of the $H_2O$ and air admixtures, which are emitted from the walls of the chamber, is gradually increasing to about 100 ppm during the first hour of operation. It is important to mention that these admixtures are working as quenching admixtures, which makes possible to obtain large multiplication coefficients of the MGEM.

<u>Investigations of the influence of $H_2O$ and $N_2$ micro admixtures to the operation of MGEM</u>

To remove the air and water from the walls of the MGEM chamber, it was baked at 200°C and pumped during 2 hours. After that, the pure commercial neon was used to fill the chamber. During several hours the concentrations of admixtures $H_2O$ and $N_2$ in neon fill of the chamber stayed at the level of several ppm, which means the neon was practically pure.

The shapes of signals of the MGEM, taken for different content of gas fills, are shown in Table I.

When the MGEM chamber is filled with pure (commercial) neon, the absence of necessary quantity of quenching admixtures results in a large photo effect from photons emitted by avalanches. This restrict the maximum coefficient of electron multiplication of MGEM to $K_{ampl}^{\max} \cong 10$ for the gas pressure of P=1 bar and to $K_{ampl}^{\max} \cong 60$ for P=0.4 bar, while the $\beta$-irradiation is used for the ionization of the gas. These results are in a good agreement with previous works [4, 5].

The proportional signals in pure neon for the pressures P=0.4 bar and 1.0 bar have a rise time, caused by primary avalanches, of $\Delta t \cong 30 \mu s$ and a tail, caused by secondary avalanches, of $\Delta t \cong 70 \mu s$. For the threshold potential difference $V=V(K_{ampl}^{\max}) \pm 2$ V, which corresponds to the maximal amplification of the MGEM, the proportional signals change into streamers with duration of ~5 ms. At further increase of the voltage by several volts, the streamers transform into continuous discharge.

For the mixture *Ne+100 ppm* $H_2O$, P=1 bar and $\beta$-irradiation, the secondary avalanche is delayed relatively to the primary one. When the voltage is increased to maximum, the amplitudes of the secondary avalanches become several times larger compared to the primary avalanches. The admixture of $H_2O$ in neon decreases the probability of formation of the steamers even at the $K^{\max}_{ampl}$. It is necessary to note that the electrons from MGEM are effectively extracted in the anode gap only at the voltage in anode gap of 470 Volts, while the corresponding multiplication in the gap is equal to $K^{anode}_{ampl}$=66. The reason for such a bad permeability of the lower electrode of the MGEM consists in a low ratio between the diameter of the openings and the thickness of the electrodes (1:1).

The admixture of $N_2$ to neon is a bad quenching addition, because of a large photo effect in mixtures $Ne+N_2$. The reason of that consists in presence of available metastable states of $N_2$ molecules with energies of 6.2 eV (1.3-2.6 s) and 8.4 eV (0.5 s).

The best results were obtained for mixtures *Ne+(1-12) ppm* $H_2O$*+(10-100) ppm* $N_2$*, 1 bar*. In the range of proportional amplification, the signals with rise time (60-80) µs for these mixtures are observed, and at threshold voltage V ($K^{\max}_{ampl}$) the streamers with broken form of duration of (5-10) ms are observed.

## Conclusion

The design of multi channel gas electron multiplier with solid metallic electrodes was proposed, produced and tested. The electrodes of the MGEM are produced from the brass plates with thickness of 1 mm, with openings of 1 mm in diameter and 1.5 steps between them. The gap between the electrodes is equal to 3 mm, and the total working area has a diameter of 20 mm.

For chamber filled with *Ne+((*$H_2O+N_2$*)≅100 ppm), P=1 bar*, and $\beta$-irradiation, the total coefficients of electron multiplication were obtained equal to (Fig. 3):

$K^{tot}_{ampl}(\beta) = K^{\max}_{ampl}(MGEM) \times K^{anode}_{ampl} = 3\cdot10^3\times10=3\cdot10^4$,

and for $\alpha$-irradiation:

$K^{tot}_{ampl}(\alpha) = K^{\max}_{ampl}(MGEM) \times K^{anode}_{ampl} = 1.9\cdot10^3\times10=1.2\cdot10^4$,

For chamber filled with *Ne+((*$H_2O+N_2$*)≅100 ppm), P=0.4 bar* (Fig. 4):

$K^{tot}_{ampl}(\beta) = K^{\max}_{ampl}(MGEM) \times K^{anode}_{ampl} = 1.33\cdot10^3\times15=2\cdot10^5$,

and $K_{ampl}^{tot}(\alpha) = K_{ampl}^{\max}(MGEM) \times K_{ampl}^{anode} = 3\cdot10^3\times15=4.5\cdot10^4$.

The necessity of use a large anode voltage is explained by low permeability of the lower electrode of MGEM. At large anode voltage the electrons are multiplied in the anode gap. For improvement of the lower electrode permeability it is necessary to increase the opening diameter of MGEM up to 1.5-2.0 mm and to decrease the electrodes thickness down to ~0.5 mm.

It is shown that the best quenching mixture for filling the chamber is *Ne+(1-12) ppm* $H_2O$*+(10-100) ppm* $N_2$, as compared with mixtures $Ne+H_2O$ and $Ne+N_2$.

## References.

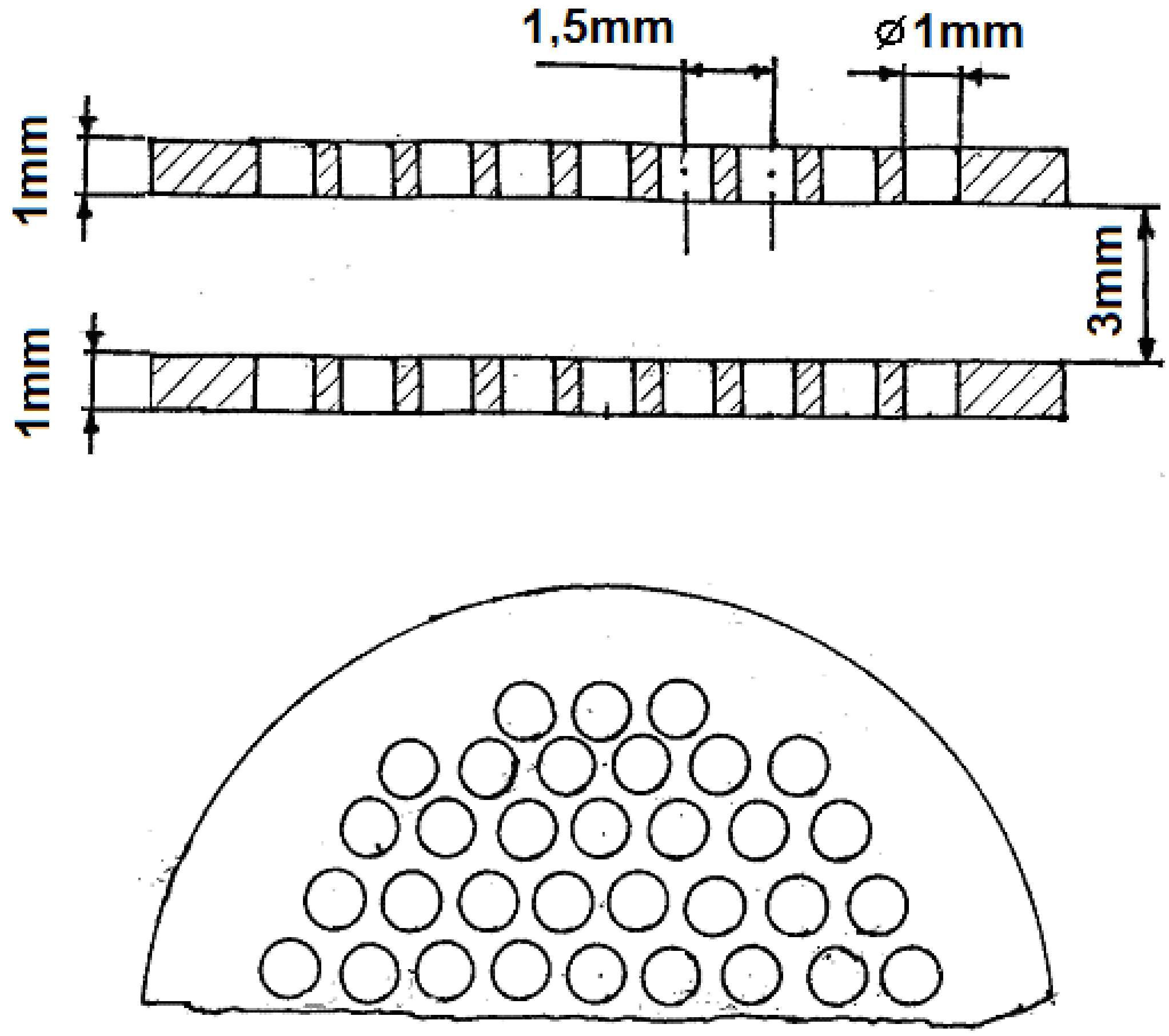

Fig.1. MGEM detector layout.

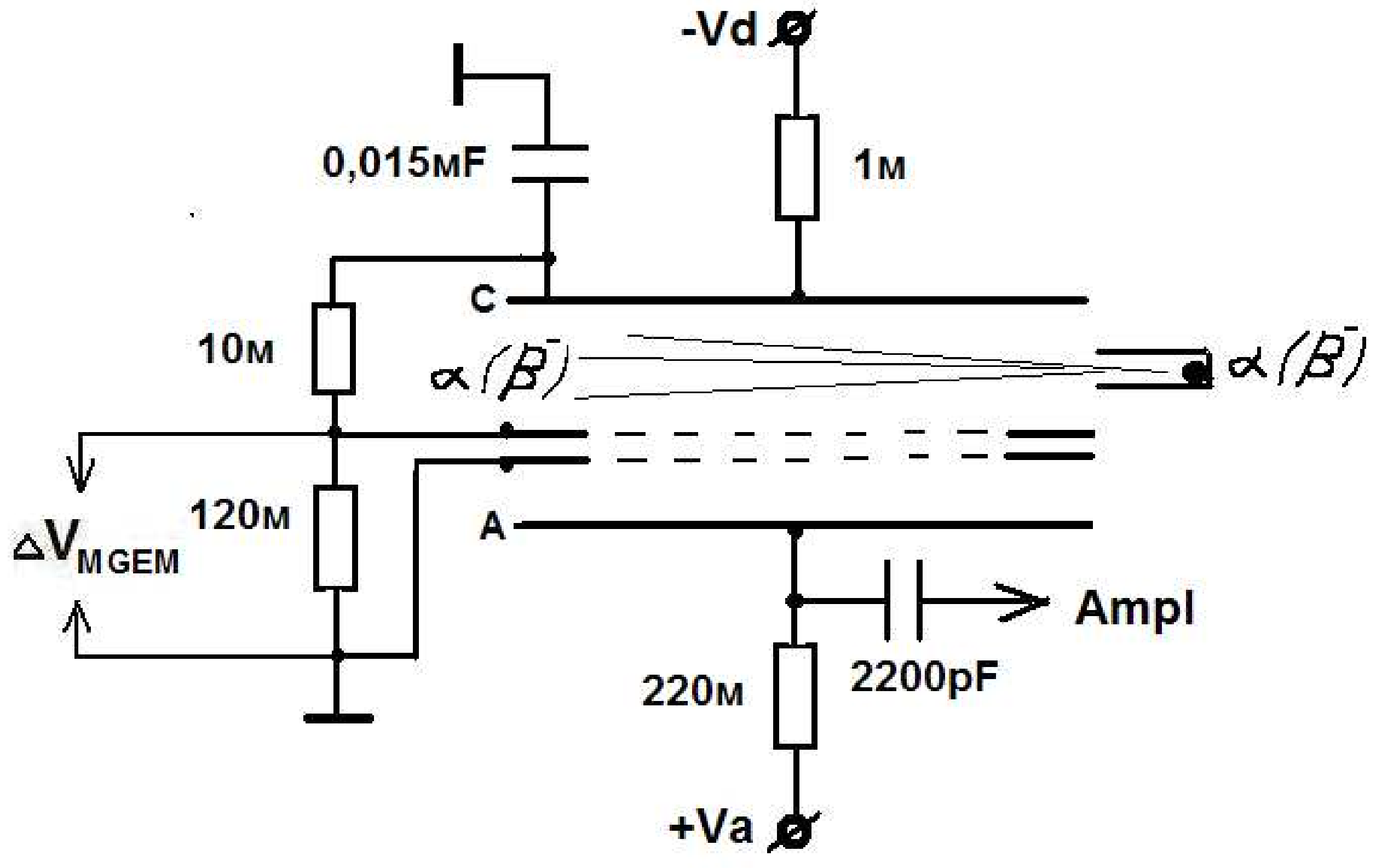

Fig.2. The chamber for MGEM tests.

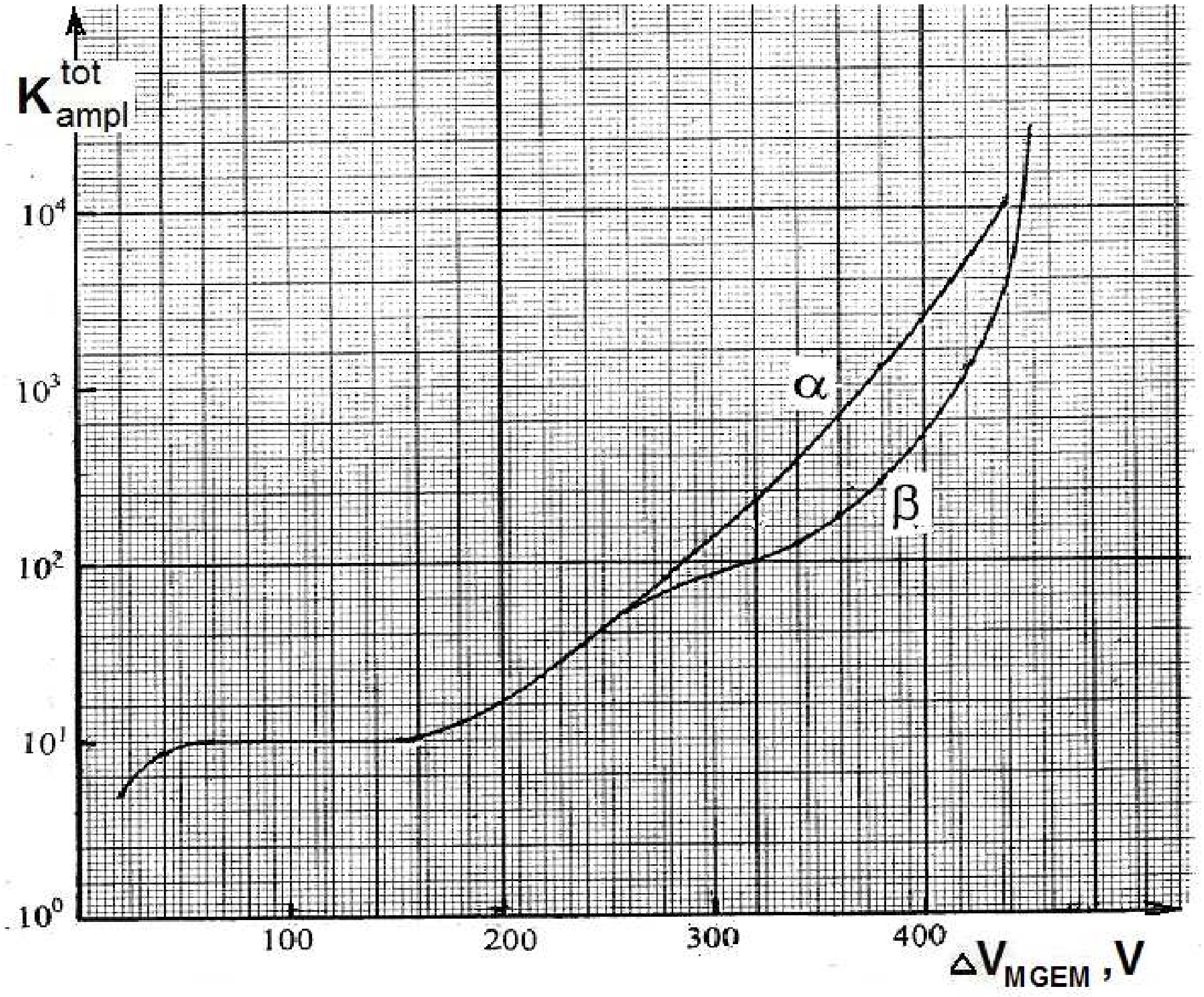

Fig.3. Dependences for the total proportional electron multiplication coefficients $K_{ampl}^{tot}$ from the potential difference between MGEM electrodes at pressure of 1 bar and for the amplification of electrons in anode gap ( $K_{ampl}^{anode}$ =10).

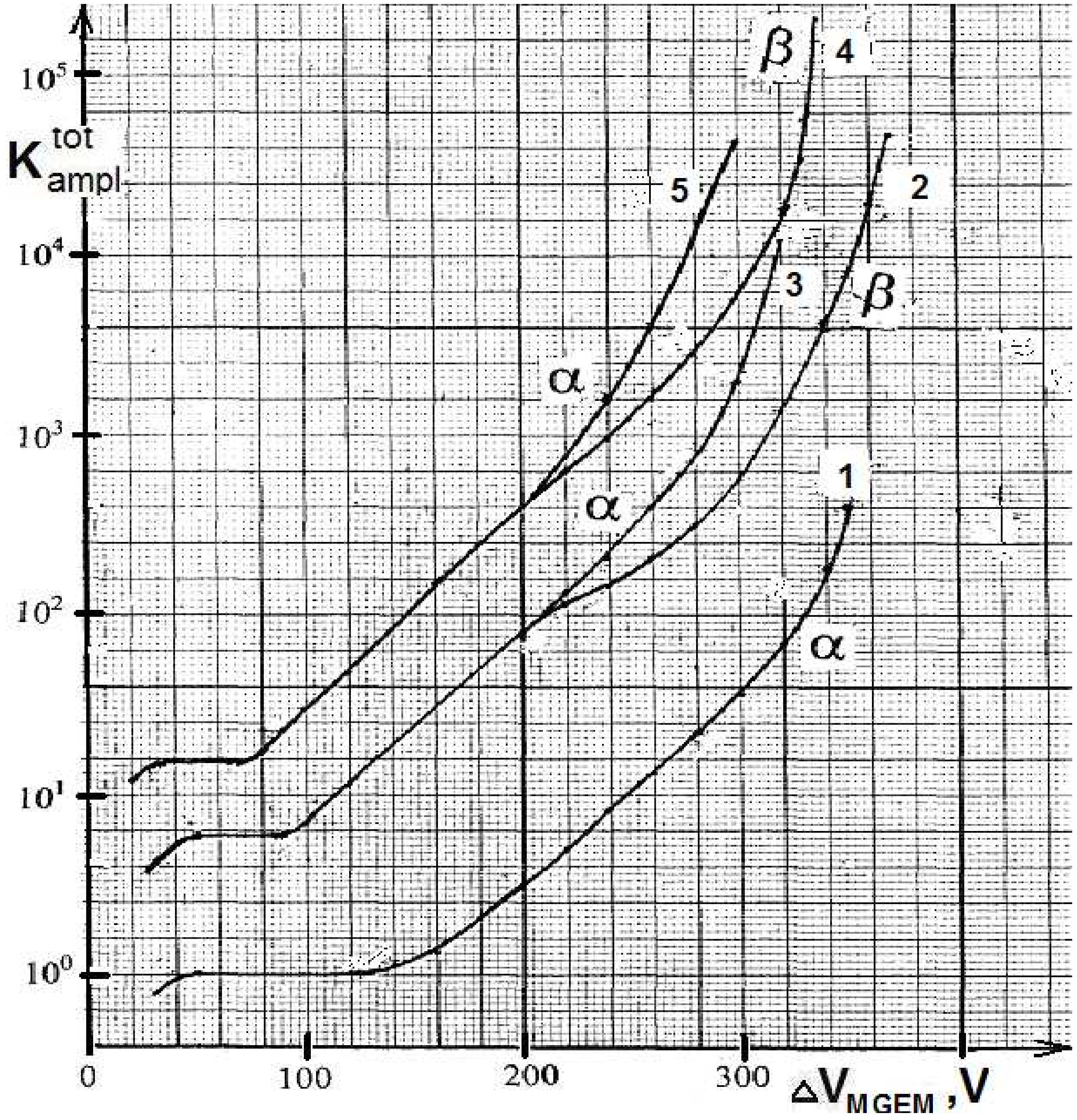

Fig.4. Dependences for the total proportional electron multiplication coefficients $K_{ampl}^{tot}$ from the potential difference between MGEM electrodes at pressure of 0.4 bar and for the amplification of electrons in anode gap:

Curve 1 corresponds to $V_a$=*250V* ($K_{ampl}^{anode}$ =1);

Curves 2,3 correspond to $V_a$=*300V* ($K_{ampl}^{anode}$ =6);

Curves 4,5 correspond to $V_a$=*320V* ($K_{ampl}^{anode}$ =15).

Table 1. The signal form for different fillings of chamber at β-irradiation

| **Gas content** | P (atm) | $+V_A$ (V) | $K_{ampl}^{anode}$ | $K_{ampl}^{max}$ | Signal form at proportional region | Signal form at maximal amplification |
|---|---|---|---|---|---|---|
| Pure *Ne* | 1.0 | 400 | 6 | 10 | 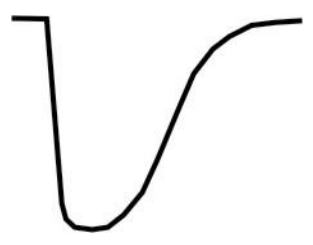 | 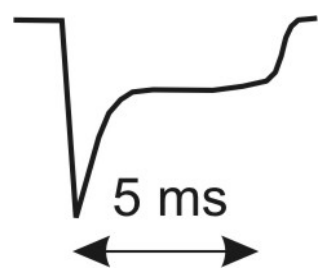  |
| Pure *Ne* | 0.4 | 250 | 1 | 60 | 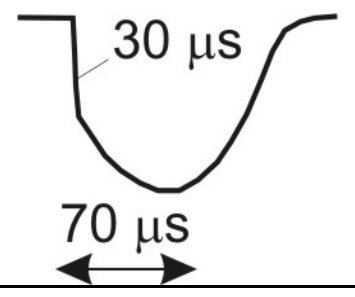  | 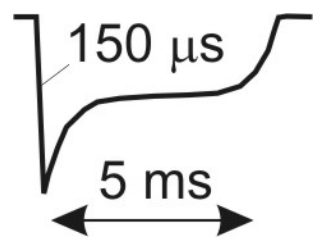  |
| *Ne + 100 ppm* $H_2O$ | 1.0 | 470 | 66 | | 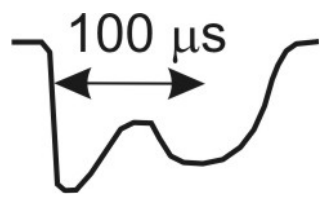  | 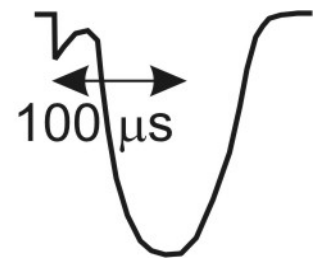  |
| *Ne + 100 ppm* $N_2$ | 1.0 | 400 | 6 | | 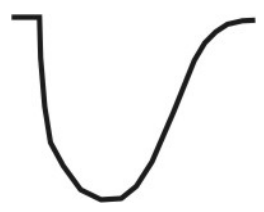 | 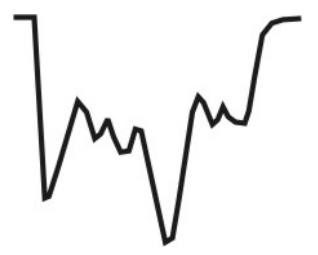 |
| *Ne + 100 ppm* $N_2$ | 0.4 | 440 | 10 | | 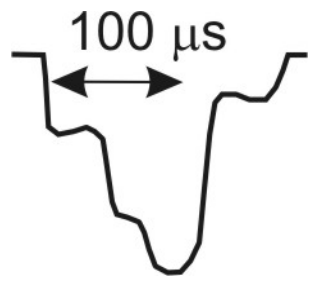  | |
| *Ne + 200 ppm* $N_2$ | 1.0 | 400 | 6 | | 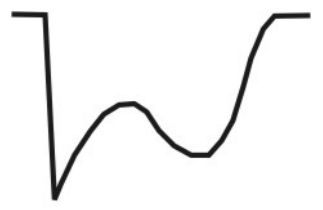 | 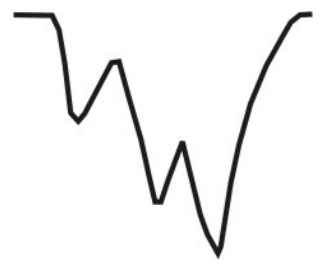 |
| *Ne + 500 ppm* $N_2$ | 1.0 | 400 | 6 | | 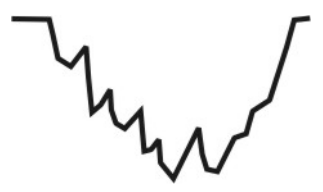 | 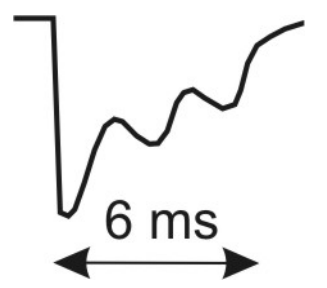  |
| *Ne + 1 ppm* $H_2O$ *+10 ppm* $N_2$ | 1.0 | 400 | 6 | | 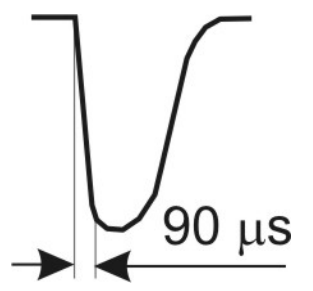  | 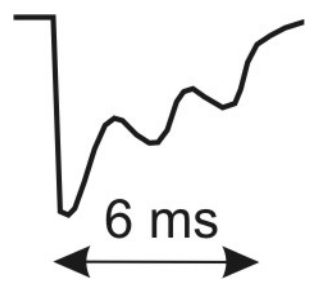  |
| *Ne + 6 ppm* $H_2O$ *+50 ppm* $N_2$ | 1.0 | 400 | 6 | | 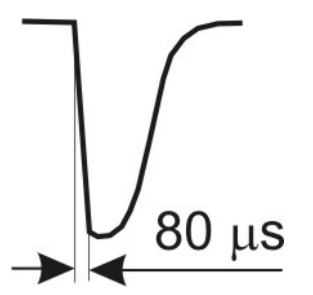  | 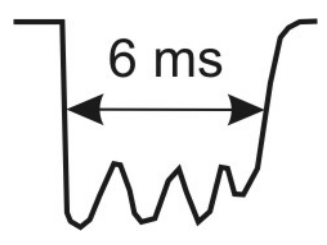  |
| *Ne + 12 ppm* $H_2O$ *+100 ppm* $N_2$ | 0.4 | 280 | 4 | | 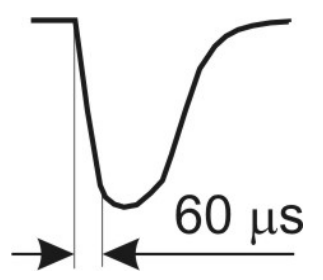  | 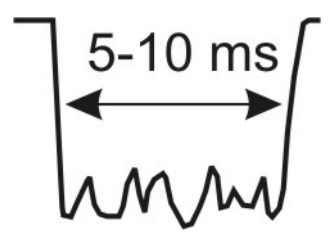  |